# Energy levels of interacting curved nano-magnets in a frustrated geometry : increasing accuracy when using finite difference methods

H. Riahi[1], F. Montaigne[1], N. Rougemaille[2], B. Canals[2], D. Lacour[1] and M. Hehn[1]

[1] Institut Jean Lamour, CNRS - Université de Lorraine, boulevard des aiguillettes
BP 70239, F-54506 Vandoeuvre lès Nancy, France
[2] Institut Néel, CNRS-UJF, BP 166, 38042 Grenoble Cedex 9, France

The accuracy of finite difference methods is related to the mesh choice and cell size. Concerning the micromagnetism of nano-objects, we show here that discretization issues can drastically affect the symmetry of the problem and therefore the resulting computed properties of lattices of interacting curved nanomagnets. In this paper, we detail these effects for the multiaxe kagome lattice. Using the Oommf finite difference method, we propose an alternative way of discretizing the nanomagnet shape via a variable moment per cell scheme. This method is shown to be efficient in reducing discretization effects.

Recently, magnetic frustration in artificial spin ice has been the subject of intense investigation and micromagnetic simulations have been used to compute and compare the energies of different spin configurations [1], to estimate the coupling coefficients between nanomagnets [2,3], to study magnetic domain wall configurations at a vertex [4, 5] or to investigate magnetization reversal processes [6]. It has also been shown that micromagnetic aspects (non uniform magnetic configurations) brings additional complexity into the physics of monopoles that is absent in spin models and that considerably enriches the physics of artificial frustrated systems. As an example, in addition to a fractionalized classical magnetic charge, monopoles in the artificial kagome are chiral [7]. However, the quantitative description of micromagnetic properties raises specific difficulties from numerical point of view, especially the choice of the mesh used for computation and the computation method. Indeed, in simulations based on a finite difference (FD) approach, i.e. the system is discretized using an orthorhombic mesh, the staircase shaped boarders linked to the meshing can significantly change the result of the simulation and induces specific effects related to the mesh and not to the system itself. The influence of the mesh increases when the system surface is curved or when the symmetry of the system is not rectangular. Therefore, beyond an inaccuracy on the calculation of the coupling coefficients or coercive fields, the effects of discretization can drastically impact the results of the computation and the FD numerical method may not be able to cover all aspects of the underlying physics.

Considering the kagome spin ice system, the degeneracy of its fundamental state is related both to its rotational symmetry ($C_3$) and its translational invariance. From numerical point of view, both aspects are impacted by discretization effects. First as the symmetry of the finite difference rectangular mesh differs from the symmetry of the vertex, the discretization of the nanomagnets along the three directions is different. Second, as the kagome lattice is incommensurable with a square discretization mesh, translated vertex in the kagome geometry will not have equivalent discretization, resulting in different computed micromagnetic configurations even for equivalent objects. As a result, the translational invariance is numerically broken. Both effects can change significantly the computation results and might change the nature of the simulated fundamental state or the simulated reversal processes. Characterizing the effects of discretization and possibly reducing their influence is therefore crucial when dealing with this kind of micromagnetic simulations. In this paper, using FD numerical method, we will propose to use a discretization scheme with variable magnetization values to reduce the consequences of meshing without increasing computation time.

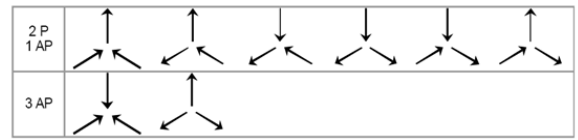

Fig.1 : The 6 degenerate low-energy configurations ("2 in / 1 out" or "2 out / 1 in" states) belonging to the ground state, and 2 degenerate high-energy configurations violating the ice-rule ("3 in" or "3 out" states).

We investigate the impact of mesh size and orientation at the scale of a vertex in the kagome lattice. This is a region where three magnetic nanomagnets are in close interaction through their stray fields and where the accuracy of the energy will strongly depend on the meshing. The vertex is made of 3 unconnected cobalt nanomagnets with dimensions $500 \times 100 \times 10$ nm$^3$. The minimum distance between nanomagnets is set to 42 nm. The micromagnetic parameters chosen are representative from polycrystalline cobalt (magnetization of $1400.10^3$ A/m and exchange of $3.10^{11}$ J/m, no crystalline anisotropy). With those parameters, magnetization lies in plane and is mostly oriented along the long axis of the nanomagnets. As a result, each nanomagnet has an Ising-like magnetic configuration. In a vertex, the number of accessible multi-axes Ising-like configurations is $2^3$ (see fig.1), 6 degenerate low-energy configurations ("2 in / 1 out" or "2 out / 1 in" states) belonging to the ground state, and 2 degenerate high-energy configurations violating the ice-rule ("3 in" or "3 out" states). Due to the effect of discretization, the three nanomagnets are not equivalently meshed and the degeneracy between low-energy configurations is lifted.

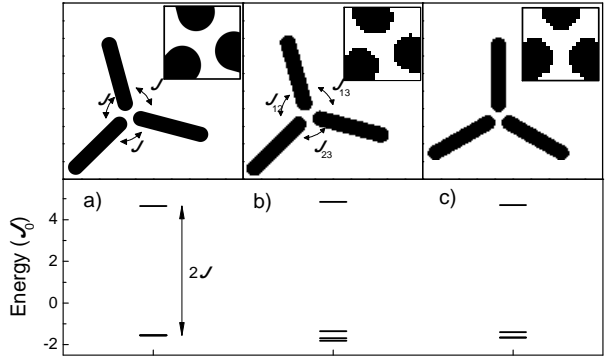

Fig.2 : Energies of the 8 different magnetic configurations at a vertex determined using (a) FE approach and with different discretization schemes using FD: (b) mesh and directions of symmetry not aligned, (c) one mesh direction aligned with one direction of symmetry of the vertex. Inset : zoom at the vertex core to show the discretization.

Most of the reported simulation results as well as the proposed method are implemented using the open source OOMMF 2D micromagnetic code [8], which is widely used by the scientific community. For this study, we compared the results obtained with this FD based code (i.e OOMMF) to results obtained with a finite elements (FE) based micromagnetic code : the FEELLGOOD software [9, 10]. In order to have a reference to compare the FD method proposed in this paper, the FE approach has been used to calculate energies of the vertex with cell size of 4nm. Combined with the more accurate description of the nanomagnets shape, the calculation results are not tainted of discretization artefacts. Figure 2 represents the energies of the 8 possible configurations calculated using the FE

FEELLGOOD software (a) and with the FD OOMMF software with two different meshes (b, c) with 10 nm cell sizes. Energies are expressed in units of $J_0$, the coupling coefficient between first neighbours within the punctual dipole approximation. A value of $J_0 = 1.34\times10^{-18}$ Joules is obtained considering the total magnetic moment of each nanomagnet concentrated at the centroid and using the vertex geometry for the distance between dipoles. The 10 nm cell size meshing induces a non-negligible degeneracy lift of the fundamental state. The comparison with the FE calculation shows that this splitting is induced by the meshing. Furthermore, the values of energy depend on the mesh (difference between (b) and (c)). This is a consequence of the mesh induced numerical roughness at the edges of the nanomagnets. The energy splitting occurs because each nanomagnet is discretized differently, depending on its position and relative angle to the mesh.

The energy of the configurations can also be seen as a sum of pair interactions (this is not generally true in the case of a micromagnetic system but still mathematically possible for a 3 nanomagnets systems). If discretization effects can be neglected and 3 folds symmetry is conserved, the energy of a vertex can be written as

$$E = -J(\vec{S}_1.\vec{S}_2 + \vec{S}_1.\vec{S}_3 + \vec{S}_2.\vec{S}_3) \qquad (1)$$

where $\vec{S}_i$ (i=1,2,3) are the unit vectors collinear to the magnetic moment of the $i$-th nanomagnet. The energy of the 6 configurations that obey the ice rule, 2 in/1 out or 1 out/2 in pseudo-spins, equals $E_I=-(1/2) J$ while the energy of the 2 configurations violating the ice rule, 3 in ou 3 out pseudo-spins, equals $E_{II}=(3/2)J$. The extraction of J is then straightforward since J=$E_{II}+E_I$. The coupling coefficient $J$ converges towards $J_0$ as the distance between nanomagnets increases. On the other hand, for smaller gaps between nanomagnets, the coupling coefficient is much larger than the one given by the punctual dipole approximation. Actually at small distances, the spatial repartition of the magnetization is not negligible through the choice of the mesh and the multipolar nature of the interaction between nanomagnets has to be taken into account.

In the more general case, i.e. with discretization effects, the energy of the vertex can be written as

$$E = -J_{12}.\vec{S}_1.\vec{S}_2 - J_{13}.\vec{S}_1.\vec{S}_3 - J_{23}.\vec{S}_2.\vec{S}_3 \qquad (2)$$

Due to time reversal symmetry, there are thus 4 levels of energy associated to the eight magnetic configurations: one for the high energy state ( $E_{II} = \frac{1}{2}(J_{12} + J_{13} + J_{23})$ ) and three for the states obeying the ice rule ( $E_I^1 = \frac{1}{2}(J_{12} - J_{13} - J_{23})$ , $E_I^2 = \frac{1}{2}(-J_{12} + J_{13} - J_{23})$ and $E_I^3 = \frac{1}{2}(-J_{12} - J_{13} + J_{23})$ ). The 4 energies associated to the eight configurations are calculated by micromagnetic simulation and therefore the three coupling coefficients can be deduced from 4 equations (an energy constant is present in the micromagnetic simulations). The spread of the coupling values comes from the mesh chosen for the calculation.

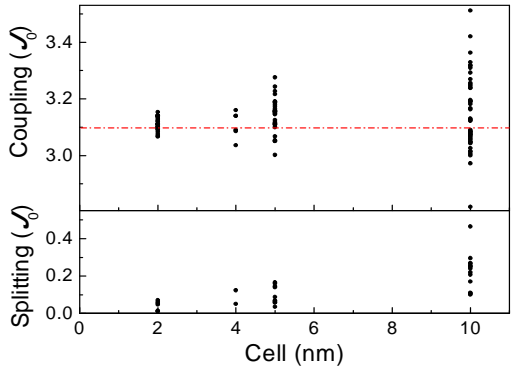

Fig.3 : Top: first neighbour coupling as a function of cell size and mesh in units of $J_0$ when using a black and white mask to define the nanostructures. Bottom: splitting of the coupling between nanomagnets as a function of cell size and mesh in units of $J_0$ when using a black and white mask to define the nanostructures. In top and bottom, the dotted line is the result of the calculation using the home made finite element FEELLGOOD software using a 4nm cell size.

In figure 3 are summarized the values of $J_{ij}$ in units of $J_0$ as a function of cell size, for different choices of the mesh (for each cell size, 10 different meshes have been used, except for 4nm cell size). The red dotted line is the result of the calculations made with the FE based FEELLGOOD software. For a fixed cell size, a distribution of coupling coefficients is observed. This distribution originates from both the asymmetry induced by the discretization for a given mesh and different realizations of the mesh. Figure 3 also represents the splitting (maximum minus minimum energy of the fundamental state).

As expected, the decrease of the cell size allows a better meshing of the nanomagnet and so a more accurate estimation of the system energy. As a result, both the spread of the $J_{ij}$ values and the splitting of the fundamental level are reduced. This conclusion is robust against varying the mesh orientation, especially rotating the mesh with respect to the vertex. A special mesh is the symmetric one as illustrated in the inset of figure 2c. As the mesh and the object shares a common symmetry axis, the fundamental state is split in only 2 levels and two coupling coefficients are equal ($J_{12} = J_{13}$ for example). Even if the number of level is reduced, the splitting between them is, in average, not reduced for a symmetric mesh. However it might be of interest for some simulations to reduce the number of different coefficients to describe the system (for the determination of coupling coefficient beyond the first neighbour [2] for example). This reduction of the number of coefficient is only possible for one vertex of the kagome lattice. Indeed, as mentioned above, due to the incommensurability between the lattice and the mesh, the number of coefficients increases rapidly with the number of simulated nanomagnets since translated vertices do not have to same mesh position with respect to the edges.

Generally, the need of minimising the discretization effects requires a small enough cell size of the mesh. As a consequence, the number of nanomagnets which can be simulated is limited due to both calculation time and memory capacities. We will now evaluate a different approach to minimise these discretization effects and increase the accuracy of the energy. In order to "smoothen" the shape of the nanomagnets, it is possible to attribute a reduced magnetic moment to cells of the mesh crossing the edges in each nanomagnet. In a conventional discretization approach, when the surface of a cell of the mesh is entirely in the nanomagnet or crosses the nanomagnet edges, the magnetic moment of the cell is the one of the bulk magnetic material; when not, magnetic moment is zero. We propose to use a grey scale mask and the grey scale value of each cell is proportional to the surface of the cell inside the nanostructure (Fig. 4).

The grey level is then made proportional to the magnetic moment carried within the pixel by modulating the thickness of the material. Figure 5 compares the results of the computed coupling coefficients $J_{ij}$ as a function of cell size and mesh in units of $J_0$ when using a black and white mask or a grey scale mask to define the nanostructures (data for column (a) are identical to those reported in figure 3). Using the grey scale mask, both the distribution of coupling coefficients and the energy splitting are drastically reduced for a given cell size. Typically the distribution obtained for a 2 nm cell size in the black and white case is obtained for a 5 nm cell size (and consequently for a large reduction of the computation time and memory occupation) by using the grey scale mask. This highlights the interest in using this kind of masks :

the same accuracy can be obtained with less computation time or a better accuracy can be obtained with same computation time.

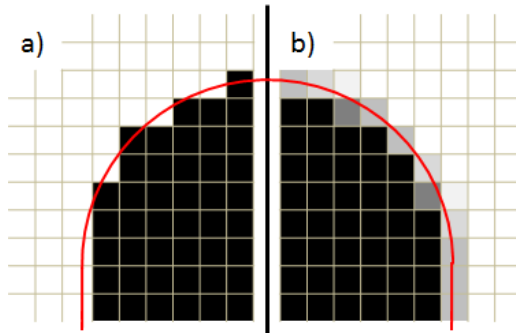

Fig.4 : The definition of the shape of a rounded nanomagnet with a square mesh (magnetic material present inside the red curve, no magnetic material outside). (a) A black and white mask defines the shape of a rounded nanomagnet. When the surface of a cell of the mesh is entirely in the nanomagnet or crosses the nanomagnet edges, the magnetic moment of the cell is the one of the bulk magnetic material; when not, magnetic moment is zero; (b) A grey scale mask defines the shape of a rounded nanomagnet. The grey scale value is proportional to the surface of the cell inside the nanostructure and is then made proportional to the magnetic moment of the cell by modulating the thickness of the material.

In this study we have investigated the impact of the mesh on the energy of interacting nanomagnets. We have shown that the choice of the mesh induces an energy splitting of the ground level of vertex and coupling coefficients that depends on the mesh size and direction. We have shown that the use of a grey scale mask allows to improve the accuracy of the energy determination and reproduces the results obtained in a FE approach. This is of prime importance in such arrays for which the energy splittings between the 6 configurations satisfying the ice rule that appear using FD are of same magnitude and even stronger than the energy splitting related to the magnetostatic interaction that lifts the degeneracy in arrays containing many vertices. Theoretically, the dipolar interaction energy with the second neighbours in the kagome geometry is equal to $0.137 \times J_0$, typically the spread in energy with the best meshing using a black and white mask with 5 nm cell size (figure 5). It is then crucial when dealing with energy determination of magnetic configurations in a kagome artificial spin system to check the origin of the calculated energy splittings. An accurate description description of the energy level is obviously necessary to study the influence of the micromagnetic aspects in the ideal system but it is also mandatory in order to introduce voluntary shape roughness or other "defects". Indeed, artificial nanomagnets are hardly perfect and do not exhibit full symmetries of the idealized lattice [11].

The authors would like to thank J.-C. Toussaint for fruitful discussions and for help with the FE FEELLGOOD calculations. This work has been partially supported by the Region Lorraine, the ANR Project FRUSTRATED and the Institut Carnot ICEEL Lorraine.

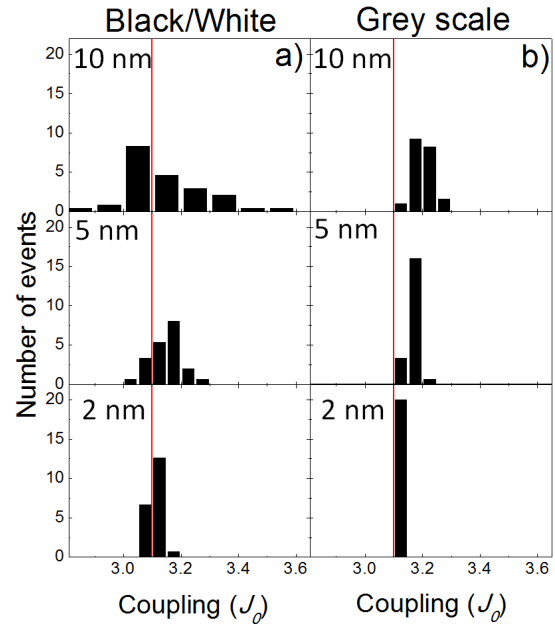

Fig.5 : Spread of the first neighbour coupling constant in units of $J_0$ as a function of cell size and mesh grid when using (a) the conventional black and white mask and (b) a grey scale mask to define the nanomagnets. The red line is the result of the calculation using the home made finite element FEELLGOOD software using a 4nm cell size.